\documentclass[prd,preprintnumbers]{revtex4}

\usepackage{amsmath}
\usepackage{amsfonts}
\usepackage{graphicx}

\begin{document}
\tolerance=5000
\def\pp{{\, \mid \hskip -1.5mm =}}
\def\cL{{\cal L}}
\def\be{\begin{equation}}
\def\ee{\end{equation}}
\def\bea{\begin{eqnarray}}
\def\eea{\end{eqnarray}}
\def\tr{{\rm tr}\, }
\def\nn{\nonumber \\}
\def\e{{\rm e}}

\preprint{}

\title{Dark energy generated from a (super)string effective action \\
with higher order curvature corrections and a dynamical dilaton}

\author{E. Elizalde$^{1}$, S. Jhingan$^{2}$, S. Nojiri$^{3}$,
S. D. Odintsov$^{4}$, M. Sami$^{2}$ and I. Thongkool$^{2}$}

\affiliation{$^{1}$Consejo Superior de Investigaciones Cient\'\i
ficas ICE/CSIC-IEEC, Campus UAB, Facultat de Ci\`encies, Torre
C5-Parell-2a pl, E-08193 Bellaterra (Barcelona) Spain}

\affiliation{$^{2}$Center of Theoretical Physics, Jamia Millia
Islamia, Jamia Nagar, Delhi-110092, India}

\affiliation{$^{3}$Department of Physics, Nagoya University, Nagoya
464-8602. Japan}

\affiliation{$^{4}$Instituci\`{o} Catalana de Recerca i Estudis Avan\c{c}ats
(ICREA) and Institut de Ciencies de l'Espai (IEEC-CSIC), Campus UAB,
Facultat de Ci\`encies, Torre C5-Par-2a pl, E-08193 Bellaterra
(Barcelona) Spain}

\date{{\small \today}}

%\date{\today}

\begin{abstract}
We investigate the possibility of a dark energy universe emerging
from an action with higher-order string loop corrections to Einstein
gravity in the presence of a massless dilaton. These curvature
corrections (up to $R^4$ order) are different depending upon the
type of (super)string model which is considered. We find in fact
that Type II,  heterotic, and bosonic strings respond differently
to dark energy. A dark energy solution is shown to exist in the
case of the bosonic string, while the other two theories do not lead
to  realistic dark energy universes. Detailed analysis of the
dynamical stability of the de-Sitter solution is presented for the
case of a bosonic string. A general prescription for the
construction of a de-Sitter solution for the low-energy
(super)string effective action is also indicated. Beyond the
low-energy (super)string effective action, when the higher-curvature
correction coefficients depend on the dilaton, the reconstruction of
the theory from the universe expansion history is done
 with a corresponding prescription for the scalar potentials.
\end{abstract}

%\pacs{}

\maketitle

%%%%%%%%%%%%%%%%%%%%%%%%%%%

\section{Introduction}
Einstein equations in their original form, with an energy-momentum
tensor for standard matter on the right hand side, cannot account
for the observed accelerated expansion of our universe. The late
time acceleration of the universe, which is directly supported by
supernovae observations, and also indirectly, through observations
of the microwave background, of the large scale structure, weak
lensing, and baryon oscillations, poses one of the most important
challenges to modern cosmology. The standard lore aimed at capturing
this important effect is related to the introduction of the
energy-momentum tensor of an exotic matter with large negative
pressure ({\it dark energy}) in the Einstein equations. The simplest
known example of dark energy (for recent reviews, see \cite{review,
review1}) is provided by the cosmological constant. This does not
require any {\it adhoc} assumption for its introduction, as is
automatically present in the Einstein equations, by virtue of the
Bianchi identities. The field theoretic understanding of $\Lambda$
is far from being satisfactory. Efforts have recently been made to
obtain $\Lambda$ in the framework of string theory, what leads to a
complicated landscape of de-Sitter vacua. It is hard to believe that
we happen to live in one of the $10^{100}$ or more vacua predicted
by the theory\cite{KKLT}. One might take the simplified view that,
like $G$, the cosmological constant $\Lambda$ is a fundamental
constant of the classical general theory of relativity and that it
should be determined from large scale observations. It is
interesting to remark that the $\Lambda CDM$ model is consistent
with the observations at present. Unfortunately, the non-evolving
nature of $\Lambda$ leads to a non-acceptable fine-tuning problem.
We do not know how the present scale of the cosmological constant is
related to Planck's or the supersymmetry breaking scale; perhaps,
some deep physics is at play here that escapes our present
understanding.

The fine-tuning problem, associated with $\Lambda$, can be
alleviated in scalar field models which do not disturb the thermal
history of the universe and can successfully mimic $\Lambda$ at late
times. A variety of scalar fields have been investigated to this
end\cite{review,Tac05,Tac105}; some of them are motivated by
field/string theory and the others are introduced owing to
phenomenological considerations. It is quite disappointing that a
scalar field description lacks predictive power; given {\it a
priori} a cosmic evolution, one can always construct a field
potential that would give rise to it. These models should, however,
not be written off, and should be judged by the generic features
which might arise from them. For instance, the tracker models have
remarkable features allowing them to alleviate the fine-tuning and
coincidence problems. Present data are insufficient in order to
conclude whether or not the dark energy has dynamics; thus, the
quest for the metamorphosis of dark energy continues.

The other alternative for getting accelerated expansion is related
to modifications of the geometry itself or the left hand side of the
Einstein equations. There are several ways of modifying gravity (for
a review, see \cite{review2}). Higher dimensional (including
stringy) effects  might lead to large-scale modifications of
gravity. Another approach, which is largely motivated by
phenomenological considerations, is related to the modification of
the form of the gravitational action (like $F(R)$ gravity, etc). The
third intriguing alternative is provided by the higher order
curvature corrections to Einstein gravity due to low-energy
(super)string effective action \cite{bento}. The leading order
correction in the string expansion parameter $\alpha'$ is given by a
Gauss-Bonnet term which has several remarkable features and which
was proposed as a dark energy model \cite{sasaki}. The
next-to-leading corrections, cubic and quartic in the curvature,
crucially depend upon the type of string model under consideration.
The higher-order curvature invariants are coupled to scalar
(dilaton/modulus) fields. One might try to fix these fields by
invoking some non-perturbative mechanism. In that case, the GB term
does not contribute to the four-dimensional equations of motion.
However, the higher order curvature terms contribute in this case.
Their presence crucially modifies the fate of a phantom dark energy
universe. Since it is difficult to realize any scenario with a fixed
dilaton or modulus field, the analysis involving dynamically
evolving fields becomes very important. A number of
papers\cite{sasaki, shinji,shinji2,mota,neupane,nos,zerbini} (see
also Ref.\cite{calcagni}) are devoted to the possibility of having
dark energy with a GB term and a dilaton/modulus field with
non-trivial potential. A steep exponential potential exhibits a
scaling solution which mimics the background (matter/radiation); the
nucleosynthesis constraint is satisfied provided the slope of the
potential is large\cite{sami}. The scaling solution describes a
decelerating universe. It is surprising that the GB term can cause a
transition from the matter dominated era to a dark energy universe
and it can also lead to transient phantom energy, provided the slope
of the exponential potential and the dilaton coupling to the GB
invariant are chosen properly \cite{sami}, or a more complicated
choice of scalar potentials is done \cite{cognola}. In such a
scenario with the exponential potential, it is quite difficult to
satisfy the nucleosynthesis constraint and, secondly, the coupling
also becomes very large. Since the introduction of the dilaton
potential needs assumptions about some non-perturbative mechanisms
and the massless dilaton naturally arises in the string loop
expansion of the low energy effective theory, it is important to
explore the possibility of a dark energy solution with a massless
dilaton. To this effect, the second order curvature correction was
considered in Ref.~\cite{nos}. This, next-to-leading correction
contains a higher order Euler density which identically vanishes for
space-time dimensions less than six; the other remaining term is a
curvature invariant of order three. The model can lead to a stable
dark energy solution. It is interesting to note that the third order
correction in $\alpha'$ crucially depends on the type of string
theory model. In this paper we incorporate string loop corrections
up to order three in $\alpha'$ to the Einstein-Hilbert action with a
massless dilaton. We investigate the cosmological dynamics of the
model and explore whether a particular string type is actually
sensitive to the existence of dark energy. We also outline a general
prescription of the construction of the de-Sitter solution in
presence of higher order curvature invariants coupled to the dilaton
field.

The paper is organized as follows. In section II, we set up the
general evolution equations from string effective Lagrangian which
incorporates  curvature corrections, up to order four in $R$,
coupled to a dynamically evolving massless dilaton. In section III,
we explore the viability of a dark energy solution for models based
upon type II, heterotic and bosonic strings in the framework of a
perturbative string theoretic set up. Section IV is devoted to the
stability analysis of the de-Sitter solution in the case of a
bosonic string model. In section V, we present a reconstruction
program for a general action with higher-order curvature invariants
coupled to dilaton functions. Section VI outlines the construction
of the de-Sitter solution in a general case. Section VII presents
our conclusions and an outlook.

%%%%%%%%%%%%%%%%%%%%%%%
\section{Evolution Equations}
The process of compactification of the string theory from higher to
four dimensions introduces scalar (moduli/dilaton) fields which are
coupled to curvature invariants. For simplicity, we shall neglect
the moduli fields associated with the radii of the internal space.
In what follows, we consider the low-energy effective string theory
action \cite{bento,shinji}
\begin{eqnarray}
\label{eq:action} {\cal S} = \int d^D x
\sqrt{-g}\left[\frac{R}{2}+{\cal L_{\phi}}+ {\cal
L}_{c}+\ldots\right]\,,
\end{eqnarray}
where $\phi$ denotes the dilaton field which is related to the
string coupling, $R$ is the scalar curvature, ${\cal L_{\phi}}$
denotes the scalar field Lagrangian, and $\mathcal{L}_c$ encodes the
string curvature correction term to the Einstein-Hilbert action
\cite{bento}
\begin{eqnarray}
&&{\cal L_{\phi}}=-\partial_{\mu}\phi\partial^{\mu}\phi-V(\phi) \,, \\
\label{stcor}
 && \mathcal{L}_c =
c_1\alpha'e^{2\frac{\phi}{\phi_{_0}}}{\cal L}_c^{(1)}+
c_2\alpha'^2e^{4\frac{\phi}{\phi_{_0}}}{\cal L}_c^{(2)}+
c_3\alpha'^3e^{6\frac{\phi}{\phi_{_0}}}{\cal L}_c^{(3)},
\end{eqnarray}
where $\alpha'$ is the string expansion parameter, ${\cal
L}_c^{(1)}$, ${\cal L}_c^{(2)}$, and ${\cal L}_c^{(3)}$ describe the
leading order (Gauss-Bonnet (GB) term), the second order and third
order curvature corrections, respectively.
The terms ${\cal L}_c^{(1)}$, ${\cal L}_c^{(2)}$ and ${\cal L}_c^{(3)}$ 
in the Lagrangian have the following form
\begin{eqnarray}
& & {\cal L}_c^{(1)}=\Omega_2\,, \\
& & {\cal L}_c^{(2)} = 2 \Omega_3 + R^{\mu\nu}_{\alpha \beta}
R^{\alpha\beta}_{\lambda\rho}
R^{\lambda\rho}_{\mu\nu}\,, \\
\label{ccc}
& & {\cal L}_c^{(3)}= {\cal L}_{31} -\delta_{H} {\cal L}_{32}
-\frac{\delta_{B}}{2}{\cal L}_{33}\,,
\end{eqnarray}
Here $\delta_B, \delta_H=0,1$ and 
\begin{eqnarray}
\hspace*{-1.0em}\Omega_2 &=& R^2-4R_{\mu \nu}R^{\mu \nu}+
R_{\mu \nu \alpha \beta}R^{\mu \nu \alpha \beta}, \\
\hspace*{-1.0em} \Omega_3 &\propto& \epsilon^{\mu \nu \rho \sigma
\tau \eta} \epsilon_{\mu '\nu '\rho '\sigma ' \tau '\eta '}
R_{\mu\nu}^{\ \ \ \mu'\nu'} R_{\rho\sigma}^{\ \ \ \rho'\sigma'}
R_{\tau\eta}^{\ \ \ \tau'\eta'}\\
\hspace*{-1.0em}  {\cal L}_{31} &=& \zeta(3)
R_{\mu\nu\rho\sigma}R^{\alpha\nu\rho\beta}\left(
R^{\mu\gamma}_{\delta\beta} R_{\alpha\gamma}^{\delta\sigma} - 2
R^{\mu\gamma}_{\delta\alpha}
R_{\beta\gamma}^{\delta\sigma} \right), \\
\hspace*{-1.0em} {\cal L}_{32} &=& \frac{1}{8} \left(
R_{\mu\nu\alpha\beta} R^{\mu\nu\alpha\beta}\right)^2
 +\frac{1}{4}  R_{\mu\nu}^{\gamma\delta}
R_{\gamma\delta}^{\rho\sigma}
R_{\rho\sigma}^{\alpha\beta}
R_{\alpha\beta}^{\mu\nu} - \frac{1}{2} R_{\mu\nu}^{\alpha\beta}
R_{\alpha\beta}^{\rho\sigma} R^\mu_{\sigma\gamma\delta}
R_\rho^{\nu\gamma\delta} - R_{\mu\nu}^{\alpha\beta}
R_{\alpha\beta}^{\rho\nu} R_{\rho\sigma}^{\gamma\delta}
R_{\gamma\delta}^{\mu\sigma}, \\
\hspace*{-1.0em}  {\cal L}_{33} &=& \left(
R_{\mu\nu\alpha\beta}R^{\mu\nu\alpha\beta}\right)^2 - 10
R_{\mu\nu\alpha\beta} R^{\mu\nu\alpha\sigma}
R_{\sigma\gamma\delta\rho}
R^{\beta\gamma\delta\rho} 
 - R_{\mu\nu\alpha\beta}
R^{\mu\nu\rho}_{\sigma}
R^{\beta\sigma\gamma\delta}
R_{\delta\gamma\rho}^{\alpha}\,.
\end{eqnarray}
The correction terms are different depending on the type of string
theory; the dependance is encoded in the curvature invariants and in
the coefficients $(c_1,c_2,c_3)$ and $\delta_H$, $\delta_B$:
\begin{itemize}
\item For the type II superstring theory: $(c_1,c_2,c_3) = (0,0,1/8)$
and $\delta_H=\delta_B=0$
\item For  the heterotic superstring theory: $(c_1,c_2,c_3) = (1/8,0,1/8)$
and $\delta_H=1,\delta_B=0$
\item For the bosonic superstring theory: $(c_1,c_2,c_3) = (1/4,1/48,1/8)$
and $\delta_H=0,\delta_B=1$
\end{itemize}

The higher order curvature corrections look complicated, in general.
However, the analysis become tractable in the case of a
Friedmann-Robertson-Walker universe. In what follows we will
specialize to the case of the FRW metric with a lapse function
$N(t)$, namely,
\begin{equation}\label{metric}
ds^2=-N(t)^2dt^2+a(t)^2\sum_{i=1}^{3} (dx^i)^2\,,
\end{equation}
 In the FRW background, the leading and next to leading corrections
simplify; they depend upon the lapse function and its time
derivative  $\dot{N}$. Since only terms linear in $\dot{N}$
contribute to the evolution equations, we shall omit the higher
powers of the time derivative of the lapse function. We then have
the following expressions for the curvature invariants
\begin{eqnarray}
\label{L2}
{\cal L}_c^{(1)}&=&\frac{24}{N^4}H^2I-\frac{24\dot{N}}
{N^6}H^3\,, \\
\label{L3}
{\cal L}_c^{(2)}&=&\frac{24}{N^6}(H^6+I^3)-\frac{72\dot{N}} {N^7}HI^2\,,
\end{eqnarray}
and
\begin{eqnarray}
\label{L41}
{\cal L}_{31} &=& -\frac{6\zeta(3)}{N^8}
\left(3H^8+4H^4I^2+4H^2I^3+I^4\right) \nonumber \\
& & +\frac{6\zeta(3)\dot{N}}{N^9}
\left(8H^5I+12H^3I^2+4HI^3\right)\,, \\
\label{L42}
{\cal L}_{32} &=& -\frac{6}{N^8}
\left(5H^8+2H^4I^2+5I^4\right)  \nonumber \\
& &+\frac{6\dot{N}}{N^9}
\left(4H^5I+20HI^3\right)\,, \\
\label{L43} {\cal L}_{33}&=& -\frac{6}{N^8} \left(60H^8+32H^4I^2 + 60I^4 \right)
\nonumber \\
& & +\frac{6 \dot{N}}{N^9} \left(64H^5I+240HI^3\right).
\end{eqnarray}
Here $I= \dot{H}+H^2$. 

Varying the action (\ref{eq:action}) with respect to $N,\dot N$, we
find the modified Friedman equation
\begin{equation}\label{eq:Friedman}
{3H^2}= \rho_c+\rho_\phi ,
\end{equation}
where
\begin{equation}\label{eq:rhocxi}
{\rho_{c}} = \sum^3_{m=1}\left \lbrace \dot{\xi_m}(\phi)\left (
\frac{\partial {\cal L}_c^{(m)}}{\partial \dot{N}} \right )+
\xi_m(\phi)\left (\frac{d}{dt}\left ( \frac{\partial {\cal
L}_c^{(m)}}{\partial \dot{N}} \right ) + 3H\frac{\partial {\cal
L}_c^{(m)}}{\partial \dot{N}}-\frac{\partial {\cal
L}_c^{(m)}}{\partial N}-{\cal L}_c^{(m)}\right)\right \rbrace
\biggr|_{N=1},
\end{equation}
and where $\xi_1(\phi)$, $\xi_2(\phi)$, $\xi_3(\phi)$ come from the
first, second and third order correction terms, respectively, and
can be written as
\begin{equation}
\xi_m(\phi)=\alpha'^m e^{2m\frac{\phi}{\phi_{_0}}} \qquad m=1,2,3.
\end{equation}
Let us consider the scalar field equation of motion derived by
varying the action (\ref{eq:action}) keeping in mind the
perturbative string theoretic description ($V(\phi)=0$)
\begin{equation}\label{eq:fieldeqn}
\ddot{\phi}+3H\dot{\phi}-\xi_1'\mathcal{L}_{c}^{(1)} -
\xi_2'\mathcal{L}_{c}^{(2)}-\xi_3'\mathcal{L}_{c}^{(3)}=0,
\end{equation}

The evolution equations look complicated, in general (see Eqs.
(\ref{GFeqn}) and (\ref{GFfr}) in the appendix), and the analysis of
the cosmological dynamics seems to be a difficult task. We,
therefore, take a different route in the search of a dark energy
solution. We shall consider the following simple solution
\cite{sasaki} and examine its viability in the present case
\begin{eqnarray}
H&=&\frac{h_0}{t},\qquad \phi=\phi_0 \ln \frac{t}{t_1},\nonumber
\end{eqnarray}
for $h_0>0$, and
\begin{eqnarray}\label{eq:powsol}
H&=&-\frac{h_0}{t_s-t},\qquad \phi=\phi_0 \ln \frac{t_s-t}{t_1},
\end{eqnarray}
when $h_0<0$ with $t_1$ as an undetermined constant. This solution
leads to a constant EOS
\begin{equation}
w_{eff}=-1-\frac{2\dot{H}}{3H^2}=-1+\frac{2}{3h_0}
\end{equation}
which corresponds to dark energy (resp.~phantom dark energy) for
$h_0>0$ (resp.~$h_0<0$), de-Sitter solution is obtained for $h_0 \to
\infty$

We will next analyze in detail whether the evolution equations,
(\ref{GFeqn}) and (\ref{GFfr}), exhibit the given solution
(\ref{eq:powsol}) for realistic values of the constants $h_0$,
$t_1$, and $\phi_0$. By substituting (\ref{eq:powsol}) into Eqs.
(\ref{GFeqn}) and (\ref{GFfr}), we obtain the algebraic equations
\begin{equation}\label{eq:fieldf}
-\phi_0^2+3h_0 \phi_0^2+f_1(h_0)X+f_2(h_0)X^2+f_3(h_0)X^3=0
\end{equation}
and
\begin{equation}\label{eq:Friedf}
\frac{\phi_0^2}{2}-{3h_0^2} + f_4(h_0)X+f_5(h_0)X^2+f_6(h_0)X^3=0,
\end{equation}
where $X \equiv \alpha'/t_1^2$ and $f'^{s}$  are given by the
following algebraic expressions
\begin{eqnarray}
f_1(h_0)&=&\delta_{HB}(12h_0^3-12h_0^4) ,\nonumber\\
f_2(h_0)&=&\delta_B\left(2h_0^3-6h_0^4+6h_0^5-4h_0^6\right) ,\nonumber\\
f_3(h_0)&=&\zeta(3)\left
(\frac{9}{2}h_0^4-36h_0^5+99h_0^6-108h_0^7
+54h_0^8\right)\nonumber\\
&&+\delta_H\left (-\frac{45}{2}h_0^4+90h_0^5-144
h_0^6+108h_0^7-54h_0^8 \right ) \nonumber \\
&&+\delta_B\left (-135h_0^4+540h_0^5-882h_0^6+ 684h_0^7-342h_0^8
\right ), \nonumber\\
f_4(h_0)&=&\delta_{HB}\left ( -12h_0^3 \right ) ,\nonumber\\
f_5(h_0)&=&\delta_B \left (-h_0^3+3h_0^5+\frac{1}{2} h_0^6 \right
) ,\nonumber\\ f_6(h_0)&=&\zeta(3)\left (-\frac{9}{4}h_0^4+15h_0^5
- \frac{57}{2}h_0^6+9h_0^7-9h_0^8\right)\nonumber \\
&&+\delta_H\left (\frac{45}{4}h_0^4-15h_0^5-15h_0^6+36h_0^7+9h_0^8
\right )\nonumber \\ &&+\delta_B \left(\frac{135}{2}h_0^4-90h_0^5-75h_0^6+198h_0^7+57h_0^8 \right )
\end{eqnarray}
where $\delta_{HB}=0, 1/2, 1$ for type II, heterotic and bosonic
string, respectively.

In what follows we would like to analyze the validity of expressions
(\ref{eq:fieldf}) and (\ref{eq:Friedf}) for realistic values of the
constants $\phi_0$ and $t_1$, corresponding to specific values of
$h_0$ relevant to dark energy observations

\section{DARK ENERGY SOLUTION}
 We shall first examine the existence of dark energy solutions
(\ref{eq:powsol}) in general and then will specialize to particular
types of string models. We will be interested in finding out whether
dark energy can distinguish amongst the string types. The case of
the bosonic string will be of special interest.

\subsection{The general case}
We can combine Eqs. (\ref{eq:fieldf}) and (\ref{eq:Friedf}) into a
single cubic equation as
\begin{equation}\label{eq:cubica}
A_3(h_0)X^3+A_2(h_0)X^2+A_1(h_0)X-6h_0^2(1-3h_0)=0,
\end{equation}
where the coefficients of $X$ are given by
\begin{equation}\label{eq:ai3}
A_m(h_0)=f_m(h_0)+2(1-3h_0)f_{3+m}(h_0)\qquad m=1,2,3.
\end{equation}
In the case of $m \leq 3$, we always have the analytic formulae for
the roots, which will be useful for the interpretation of the
relation and the contribution from each of the correction terms to
the solution.

The positivity of $X$, the real root of the cubic equation will
impose a restriction on the possible values of $h_0$. The real
solution for $(\ref{eq:cubica})$ can be obtained from the cubic root
formula, as
\begin{equation}\label{eq:realroot}
X(h_0)=s_1(h_0)+s_2(h_0)-\frac{1}{3}\frac{A_2(h_0)}{A_3(h_0)},
\end{equation}
where
\begin{equation}
s_1(h_0)=\left [ r+(q^3+r^2)^{\frac{1}{2}} \right ]^{\frac{1}{3}}, \qquad
s_2(h_0)=\left [ r-(q^3+r^2)^{\frac{1}{2}} \right ]^{\frac{1}{3}},\nonumber
\end{equation}
\begin{eqnarray}
r(h_0)&=&\frac{3h_0^2(1-3h_0)}{A_3(h_0)}+\frac{1}{6}
\frac{A_1(h_0)A_2(h_0)}{A_3(h_0)^2}
-\frac{1}{27}\left (\frac{A_2(h_0)}{A_3(h_0)}\right )^3,\nonumber \\
q(h_0)&=&\frac{1}{3}\frac{A_1(h_0)}{A_3(h_0)}
-\frac{1}{9}\left( \frac{A_2(h_0)}{A_3(h_0)} \right )^2.
\end{eqnarray}

\subsubsection{ $X(h_0) \equiv \alpha '/t_1^2 > 0$}

The cubic equation has one real root $X$ provided $q^3+r^2>0$. We
have checked numerically that the relation $q^3+r^2>0$ is true for
all $h_0$ in the region $0.8<|h_0|$, for all three string types. Eq.
(\ref{eq:realroot}) should then be used in order to find the range
of $h_0$ such that $X(h_0)>0$ and the corresponding equation of
state parameter $w_{eff}$ be confronted with the observations.

\subsubsection{ $\Omega_c  < 1$}

Another important constraint on $h_0$ is dictated by the fact that
$\phi_0^2<1$, as the dilaton is a real scalar function. The cubic
equation (\ref{eq:cubica}) does not involve $\phi_0$; it enters into
the Friedman equation through $\rho_c$ which encodes all higher
order curvature corrections. Using the Friedman equation
(\ref{eq:Friedman}) $\&$ (\ref{eq:powsol}), we find
\begin{eqnarray}
3h_0^2&=&\rho_\phi+\rho_c \nonumber,\\
1&=&\frac{\rho_\phi}{3h_0^2} + \frac{\rho_c}{3h_0^2}\equiv \frac{\phi_0^2}{6h_0^2}+\frac{\rho_c}{3 h_0^2}\\
1&=&\Omega_\phi+\Omega_c , \nonumber\label{Friedh0}
\end{eqnarray}
where $\Omega_c$ is the dimensionless density parameter contributed
by the correction terms. The constraint   $\Omega_c<1 $ is
equivalent to $\phi_0^2>0$, otherwise the dilaton would turn complex
and this would put a bound on the possible range of $h_0$ which
should be combined with the constraint dictated by the positivity of
real root of the cubic equation (\ref{eq:cubica}). The range of
$h_0$ compatible with the two constraints should then be confronted
with the observation on the equation of state for the dark energy.
The recent analysis of the three year WMAP data combined with the
supernova legacy survey (SNLS) constraints the dark energy equation
of state parameter $w_{DE}$. At $68\%$ confidence level, the best
fit value $w_{DE}$ is given by $ w_{DE}=-1.06_{-0.08}^{+0.07}$. If
the flat prior is imposed, the parameter is constrained by
$w_{DE}=-0.97_{-0.09}^{+0.07}$, which translates into a bound on
$h_0$, as $h_0 \leq -11.11$ and $h_0 \geq 6.67$.

We next turn to the individual string models to find out their
viability as dark energies, in view of the aforesaid constraints.

\subsection{Type II string}

\begin{figure}[t]
\label{fig1}
\begin{center}
\includegraphics[width=8cm,height=6cm,angle=0]{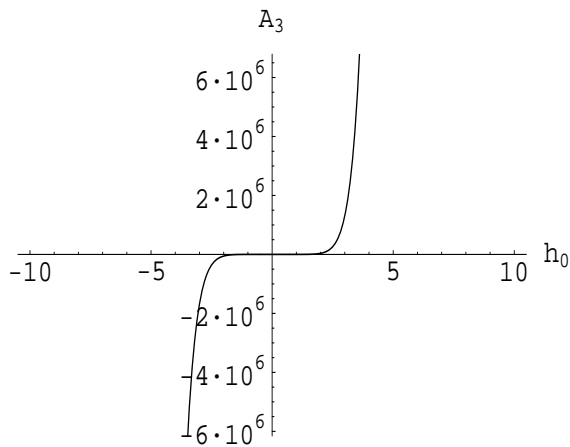}
\end{center}
 \caption{Plot of $A_3$ against $h_0$ is shown for Type
II string model.}
\end{figure}
\begin{figure}[t]
\begin{center}
\includegraphics[width=8cm,height=6cm,angle=0]{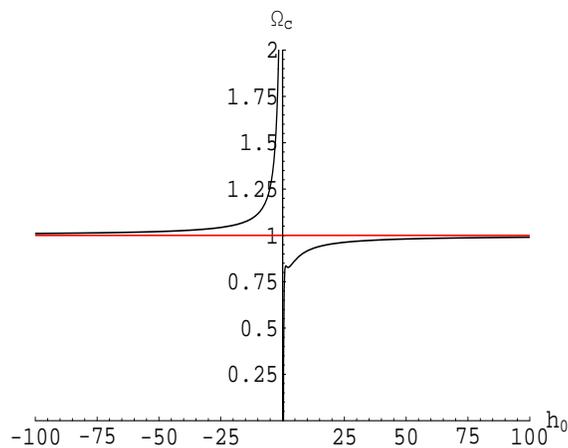}
\end{center}
\label{fig2} \caption{A plot of $\Omega_c$ against $h_0$ for the
type II string shows that $h_0>0$, the density parameter
$\Omega_c<1$.}
\end{figure}

 The case of a Type II string is simplest to investigate. In this
case, we have $A_1(h_0)=A_2(h_0)=0$ and the equation
(\ref{eq:cubica}) reduces to
\begin{equation}
A_3(h_0)X(h_0)^3-6h_0^2(1-3h_0)=0,
\end{equation}
which has the following solution
\begin{eqnarray}
X(h_0)=\left(2r(h_0) \right)^\frac{1}{3}=\left (
\frac{6h_0^2(1-3h_0)}{A_3(h_0)} \right )^\frac{1}{3}. \label{xa}
\label{X2}
\end{eqnarray}
In this case the expression of $A_3$ simplifies to
\begin{equation}
A_3(h_0)=f_3(h_0)+2(1-3h_0)f_6(h_0),
\end{equation}
with
\begin{eqnarray}
&&f_3(h_0)=\zeta(3)\left (
\frac{9}{2}h_0^4-36h_0^5+99h_0^6-108h_0^7+54h_0^8 \right ) \nonumber \\
&&f_6(h_0)=\zeta(3)\left
(-\frac{9}{4}h_0^4+15h_0^5-\frac{57}{2}h_0^6+9h_0^7-9h_0^8 \right ).
\label{f36}
\end{eqnarray}

Let us first implement the condition $X(h_0)=\alpha '/t_1^2 > 0$.
The sign of $A_3$ is important for constraining $h_0$ using the
positivity of $X$. From Eq. (\ref{f36}), we obtain
\begin{eqnarray}
A_3(h_0)=\zeta(3)\left
(\frac{15}{2}h_0^5-48h^6+81h_0^7-18h_0^8+54h_0^9 \right).\nonumber
\end{eqnarray}
We have plotted $A_3$ in Fig.~1. The plot shows that $A_3(h_0)>0$
for $h_0>0$ and $A_3(h_0)<0$ when $h_0<0$. Using Eq. (\ref{xa}), we
get the possible  region that gives $X(h_0)>0$ as $0<h_0<1/3$ which
always yields $w_{eff}>1$. Thus, no viable solution exists in this
case. Therefore, up to $4^{th}$ order corrections in $R$, the Type
II superstring model is clearly ruled out as dark energy (for the
$\Omega_c < 1$
 case, see Fig.~2).

\subsection{Heterotic string}
In this case $A_2(h_0)=0$, and $A_1(h_0)$ and $A_3(h_0)$ are given
by
\begin{eqnarray}
A_3(h_0)&=&f_3(h_0)+2(1-3h_0)f_6(h_0), \nonumber\\
A_1(h_0)&=&f_1(h_0)+2(1-3h_0)f_4(h_0), \nonumber
\end{eqnarray}
where
\begin{eqnarray}
A_{3}(h_0)&=&-\frac{15}{2}h_0^5-84h_0^6+270h_0^7-252h_0^8-54h_0^9\nonumber\\
&&+\zeta(3) \left (\frac{15}{2}h_0^5-48h_0^6+81h_0^7-18h_0^8+54h_0^9\right )\nonumber\\
&\approx&1.515h_0^5-141.699h_0^6+367.367h_0^7-273.637h_0^8+10.911h_0^9.
\end{eqnarray}
In this case, we check the consistency of $X(h_0)$ and $\Omega_c$
numerically as the formula (\ref{X2}) does not seem to hold in this
case. However, as we remark below, $A_3$ might still be used as a
yard stick for the consistency check

We plot $X(h_0)$ in Fig.~3 which shows that $X>0$ provided
$0<h_0<23.68$. This constraint should be combined with $\Omega_c
<1$. The plot of $\Omega_c$ in Fig.~4 tells us that either
$0<h_0<5.04$ or $h_0>23.68$. Thus, the allowed range for the
parameter $h_0$  is $0<h_0<5.04$, which corresponds to $w_{eff}\ge
-0.868$. Such a value of the equation of state parameter is ruled
out by recent $WMAP_3$ and SNLS survey data. However, the combined
data (CMB+LSS+SNLS) forces the equation of state to vary as $-1.001<
w_{DE}<-0.875$. This results shows that the heterotic string model
is marginally compatible with dark energy observations.

\begin{figure}[t]
\begin{center}
\includegraphics[width=8cm,height=6cm,angle=0]{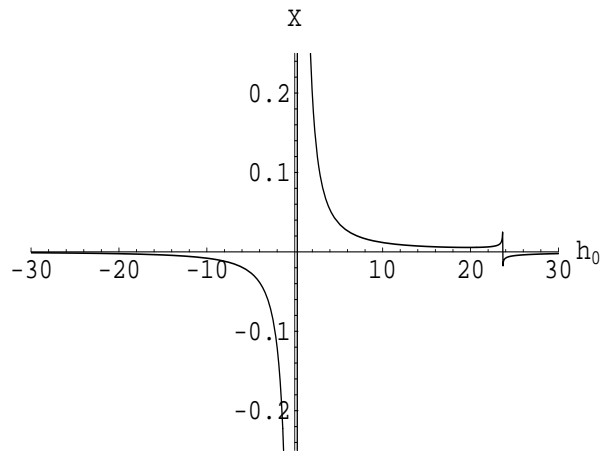}
\end{center}
\label{fig3} \caption{Plot of X against $h_0$ shows that
$0<h_0<23.68$ is the viable range of $h_0$ for heterotic string.}
\end{figure}
\begin{figure}[t]
\begin{center}
\includegraphics[width=8cm,height=6cm,angle=0]{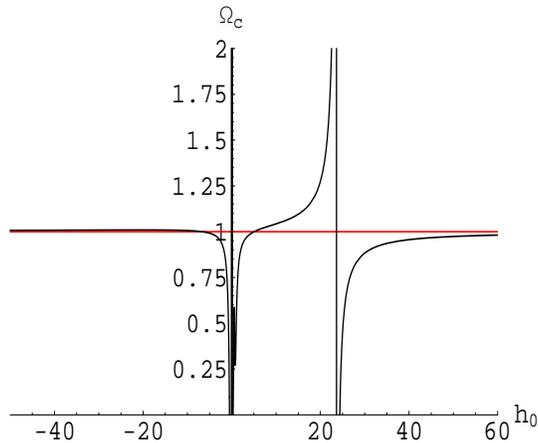}
\end{center}
\label{fig4} \caption{Plot of $\Omega_c$ versus $h_0$ for the
heterotic string model. The region given by $\Omega<1$ shows the
allowed range of $h_0$, which corresponds to $0.8 < h_0 \leq 5.04$
and $h_0 > 23.68$.}
\end{figure}

\subsection{Bosonic string}
We now turn to the bosonic string, for which $A_m \ne 0$ for
$m=1,2,3$. All $f^{s}$ contribute to $A_m$ in this case. We quote
below the expression for $A_3$.
\begin{eqnarray}
A_3(h_0)&=&-45h_0^5-492h_0^6+1530h_0^7-1416h_0^8-342h_0^9\nonumber\\
&&+\zeta(3) \left (\frac{15}{2}h_0^5-48h_0^6+81h_0^7-18h_0^8+54h_0^9\right )\nonumber\\
&\approx&-35.985h_0^5-549.699h_0^6+1627.370h_0^7-1437.640h_0^8-277.089h_0^9.
\end{eqnarray}
In order to the check if $X(h_0)>0$ and $\Omega_c <1$, we display
our numerical results in Figs.~5 and 6, which show that
\begin{itemize}
\item a. $X(h_0)>0$ for $h_0<-6.189$~or~$h_0>0 $
\item b. $\Omega_c <1$ for $h_0 >0.5$
\end{itemize}

Note that $(r^2+q^3)>0$, the condition for the existence of one real
root of (\ref{eq:cubica}), constraints $h_0$ to be $h_0>0.8$. We
therefore conclude that the allowed range for $h_0$ is given by
$h_0>0.8$ and this corresponds to $-1<w_{eff}<-0.17$. The
requirement for the dilaton to be real clearly excludes the
possibility of phantom energy. It is really interesting that the
bosonic string responds positively to the requirement of dark
energy.

Before moving to the next section, a remark about $A_3$ is in order.
In the case of a Type II superstring, $A_1$ and $A_2$ vanish
identically, leading to $s_2=0$. The sign of $A_3$ then becomes
important for the consistency check on $X(h_0)$. In the cases of
heterotic and bosonic strings this is no longer true. In these cases
we have directly checked the positivity of $X(h_0)$ using numerical
treatments. Interestingly enough, we have found numerically that,
for a generic range of the parameter $h_0$, it turns out that
$s_1>>s_2$ telling us that (\ref{X2}) still holds approximately for
numerical values of $h_0$ which are of interest to us. We then could
analyze the bosonic and heterotic models by checking the sign of
$A_3$ as we did for the case of the Type II string. In fact, our
numerical check shows that we reproduce the exact numerical results
presented here to a good accuracy.
\begin{figure}[t]
\begin{center}
\includegraphics[width=8cm,height=6cm,angle=0]{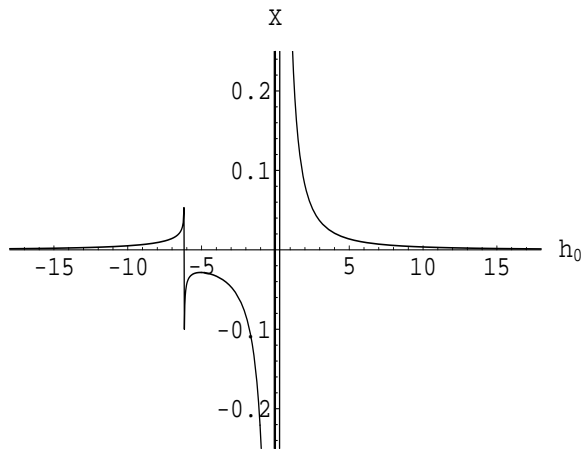}
\end{center}
\label{fig5} \caption{The plot of $X$ versus $h_0$ for the bosonic
string shows that $X(h_0)>0$ for $h_0<-6.189$~or~$h_0>0 $}.
\end{figure}

\begin{figure}[t]
\begin{center}
\includegraphics[width=8cm,height=6cm,angle=0]{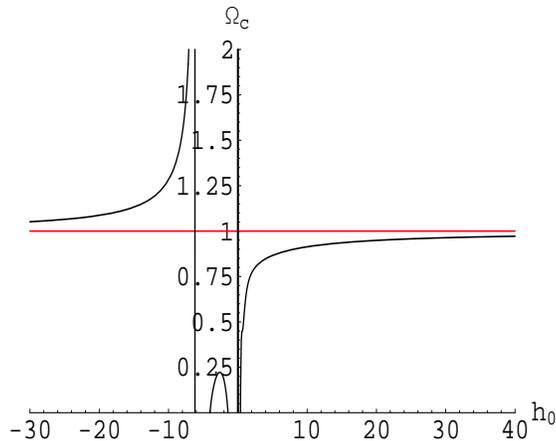}
\end{center}
\label{fig6} \caption{$\Omega_c$ is plotted here against $h_0$ for
the case of the bosonic string.}
\end{figure}

\section{The de-Sitter solution and its stability}

It is only in the bosonic case that we have the desired solution
(normal dark energy). The general expressions for the Friedman
equation and the scalar field equation of motion, given in the
appendix, are difficult to analyze in general. However, in the
de-Sitter case the equations get simplified and in what follows we
will analyze the stability of this solution.

Following Nojiri et. al \cite{sasaki} we define two new variables:
\begin{equation}\label{new-var}
X = \frac{\dot \phi}{H} , \quad Y = H^2 \alpha' e^{2\phi/\phi_0}
\end{equation}
With these new variables the Friedman equation and the equation of
motion for the scalar field (see appendix) can be written as
\begin{eqnarray}\label{new-eqns}
\frac{dX}{dN} &=& -3X +\frac{12}{\phi_0}Y + \frac{4}{\phi_0} Y^2 +
\frac{342}{\phi_0}Y^3 -\frac{54\zeta (3)}{\phi_0}Y^3 \\
\frac{dY}{dN} &=& -\frac{1}{2} +\frac{1}{12}(X^2+Y^2) +
\frac{19}{2}Y^3 - \frac{1}{\phi_0}XY^2 - \frac{144}{\phi_0}XY^3 -
{3\zeta (3)} Y^3 \left(\frac{1}{2} - \frac{6X}{\phi_0} \right)
\end{eqnarray}
These expressions are much simpler, since $H$ is a constant and its
derivatives vanish identically. For $\phi_0 = -0.01$ the fixed
points $(X_c,Y_c)$ are
\[
\begin{tabular}{|c|c|}
  \hline
  $X_c \rightarrow$  & $Y_c \rightarrow$ \\
  \hline
-25.1415 & 0.0574  \\
  -2.4860 & 0.0062  \\
   2.4860 & - 0.0062  \\
   29.4780 & -0.0680 \\
  -0.0005 $\pm $ 0.0060 $\iota$ & -0.0072 $\pm$ 0.2080 $\iota$ \\
  \hline
\end{tabular}
\]
The first critical point in the table above is relevant for us since
the second turns out to be unstable. The third and the fourth point
give a negative value to the string expansion parameter and are,
therefore, not relevant. We next examine the stability of the
solution around the critical point $(X_c,Y_c)=(-25.1415, 0.0574 )$.
The perturbation matrix ${\cal M}$ has the following form

\[
\cal M = \left(%
\begin{array}{cc}
  -3 & \frac{1}{\phi_0} (12 + 8 Y_c + [1026 - 162 \zeta (3)] Y_c^2) \\
  \frac{X_c}{6} -\frac{Y_c^2}{\phi_0}(1 + [114-18\zeta (3)]Y_c) & \frac{Y_c}{6} + \frac{Y_c^2}{2}
  [57-9\zeta (3)] - \frac{2 X_c Y_c}{\phi_0}(1 + [171 -27\zeta (3)]Y_c) \\
\end{array}%
\right)
\]

The eigen values of the stability matrix ${\cal M}$ are: $-2161.75$
and $-1.7903$. Therefore, the critical point is a stable node. In
general fixed points exist for the range $|\phi_0| \in (0,
0.05882)$.
\begin{figure}[h]
    \centering
        \includegraphics[width=8cm,height=6cm,angle=0]{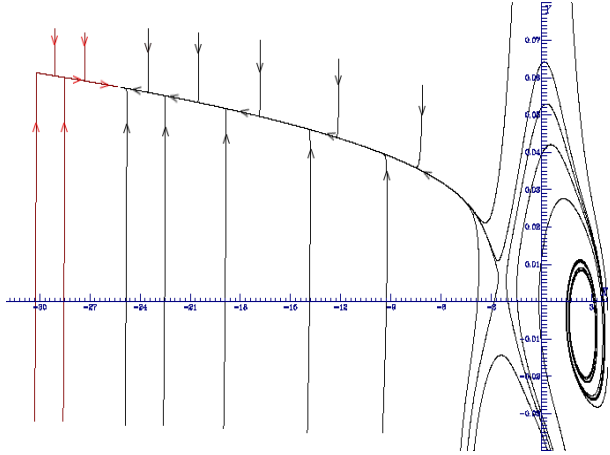}
    \label{fig:fxpts} \caption{The phase portrait of cosmological evolution
    described by (\ref{eq:action}) in case of bosonic string restricted to de-Sitter
     for $\phi_0 = -0.01$. Trajectories starting anywhere in the phase space converge at the
     stable node (-25.1415, 0.0574).}
\end{figure}

%%%%%%%%%%%%%%%%%%%%%
\section{Reconstruction of the universe expansion history: beyond the
low-energy string effective action}

In this section we study a more general gravitational action where
the coefficients of the curvature corrections depend on the dilaton.
In addition, a scalar potential is added. We then consider the
reconstruction of such general modified gravity from the universe
expansion history, following the technique developed
in\cite{reconstruct}. Let us keep $\xi_m(\phi)$ to be general
functions of the scalar field $\phi$. For simplicity, however, we
neglect ${\cal L}_c^{(3)}$. Then, the action has the following form:
\be \label{3R0} {\cal S} = \int d^4 x \sqrt{-g}\left[\frac{R}{2}
 -\partial_{\mu}\phi\partial^{\mu}\phi-V(\phi)
+ \xi_1(\phi){\cal L}_c^{(1)} + \xi_2(\phi) {\cal L}_c^{(2)}
\right]\ . \ee Even if one includes ${\cal L}_c^{(3)}$, we can
reconstruct the action, although the final expression becomes rather
complicated. Neglecting ${\cal L}_c^{(3)}$, the explicit form of the
FRW equation and the scalar field equation follows: \bea \label{3R1}
0&=&-{3}H^2 + \frac{1}{2}{\dot \phi}^2 + V(\phi) -
24\dot\xi_1(\phi)H^3
 - 72 \dot\xi_2(\phi)H\left(H^2 + \dot H\right)^2 \nn
&& + 24 \xi_2(\phi) \left\{ -6 H\left(H^2 + \dot H\right) \ddot H +
2 \left( H^2 + \dot H\right)^3
 - 18 H^2 \left( H^2 + \dot H\right)^2 + 5 H^6\right\} \ ,\\
\label{3R2} 0&=& \ddot\phi + 3H\dot\phi + V'(\phi) - 24 \xi_1'(\phi)
\left( H^2 + \dot H\right) + 24\xi_2'(\phi)\left(H^6 + \left( H^2 +
\dot H\right)^3\right)\ . \eea By combining (\ref{3R1}) and
(\ref{3R2}), we obtain \bea \label{3R3}
\xi_1(\phi(t))&=& \int dt \frac{a(t)W(t)}{H^2} \ ,\\
\label{3R4} V(\phi(t))&=& {3}H^2 - \frac{1}{2}{\dot\phi}^2 + 24 H
a(t) W(t) + 72 \dot \xi_2(\phi(t)) H \left( H^2 + \dot H\right) \nn
&& + 24 \xi_2(\phi(t))\left\{6\left( H^2 + \dot H\right) H\ddot H -
2 \left( H^2 + \dot H\right)^3
+ 18 \left( H^2 + \dot H\right)^2 H^2 - 5 H^6\right\}\ ,\\
\label{3R5} W(t)&=&\int \frac{dt}{a(t)}\left[-\frac{\dot H}{4}
 - \frac{{\dot\phi}^2}{8} +\dot\xi_2(\phi(t))\left\{ - 13H^5 - 45 H^3 \dot H
 - 3 H{\dot H}^2 + \frac{{\dot H}^3}{H} - 12
\left( H^2 + \dot H\right)\ddot H\right\} \right. \nn && \left. +
6\xi_2(\phi)\left\{ - 11H^4 \dot H - 20 H^2 {\dot H}^2
 - 4 {\dot H}^3 + \left(-5H^3 - 7H\dot H - \ddot H\right)\ddot H - \left( H^2
+ \dot H\right) \dddot H\right\}\right]\ . \eea Hence, if we
consider the theory where $\xi_1(\phi)$ and $V(\phi)$ are expressed
in terms of two functions, $g(t)$ and $f(\phi)$, and an arbitrary
$\xi_2(\phi)$ as follows (compare with reconstruction in the less
complicated case when only the first order correction is present
\cite{cognola,nos,petr}) \bea \label{3R6}
\xi_1(\phi)&=& \int dt \frac{\e^{g(f(\phi))}U(\phi)}{g'(f(\phi))^2} \ ,\\
\label{3R7} V(\phi)&=& {3}g'(f(\phi))^2 - \frac{1}{2f'(\phi)^2} + 24
g'(f(\phi)) \e^{g(f(\phi))} U(\phi) + \frac{72
\xi_2(\phi)g'(f(\phi))}{f'(\phi)} \left( g'(f(\phi))^2 +
g''(f(\phi)) \right) \nn && + 24 \xi_2(\phi)\left\{6\left(
g'(f(\phi))^2 + g''(f(\phi)) \right) g'(f(\phi)) g'''(f(\phi))
 - 2 \left( g'(f(\phi))^2 + g''(f(\phi))\right)^3 \right. \nn
&& \left. + 18 \left( g'(f(\phi))^2 + g''(f(\phi))\right)^2 g'(f(\phi))^2 - 5 g'(f(\phi))^6\right\}\ ,\\
\label{3R8} U(\phi)&=&\int
\frac{f'(\phi)d\phi}{\e^{g(f(\phi))}}\left[-\frac{g'(f(\phi))}{4}
 - \frac{1}{8f'(\phi)^2} +\frac{\xi_2'(\phi)}{f'(\phi)}\Bigl\{
 - 13g'(f(\phi))^5 - 45 g'(f(\phi))^3 g''(f(\phi)) \right. \nn && - 3
g'(f(\phi)) \left(g''(f(\phi))\right)^2 +
\frac{g''(f(\phi))^3}{g'(f(\phi))} - 12 \left(g'(f(\phi))^2 +
g''(f(\phi))\right)g'''(f(\phi)) \Bigr\} \nn && +
6\xi_2(\phi)\left\{ - 11g'(f(\phi))^4 g''(f(\phi)) - 20
g'(f(\phi))^2 g''(f(\phi))^2 - 4 g''(f(\phi))^3 \right.\nn &&
\left.\left. + \left(-5g'(f(\phi))^3 - 7g'(f(\phi))g''(f(\phi))
 - g'''(f(\phi))\right)g'''(f(\phi))
 - \left( g'(f(\phi))^2 + g''(f(\phi)) \right) g''''(f(\phi)) \right\}\right]\ .
\eea then it is not hard to check that a solution is given by \be
\label{3R9} H=g(t)\ ,\quad \phi=f^{-1}(t)\ . \ee Here $f^{-1}(t)$ is
the inverse function of $f(\phi)$.

An example of this situation is the following: \be \label{FGB12}
g(t)=H_0 t + H_1\ln\left(\frac{t}{t_0}\right)\ , \ee and $f(\phi)$
to be properly defined, we obtain \be \label{FGB13} H(t)= H_0 +
\frac{H_1}{t}\ . \ee When $t$ is small, $H$ (\ref{FGB13}) behaves as
that in a universe with a perfect fluid, with $w_{eff}=-1 + 2/3H_1$,
and when $t$ is large, $H$ behaves as in the de-Sitter space, where
$H$ is a constant. Then, if we choose $H_1=2/3$, we find that before
the acceleration epoch, the universe behaves as matter dominated one
with $w_{eff}=0$.  After that, it enters the acceleration phase.

Another example is: \be \label{FGB14} g(t)=\tilde
H_0\ln\frac{t}{t_0} - \tilde H_1\ln\left(\frac{t_0 - t}{t_0}\right)\
, \ee which gives \be \label{FGB15} H(t)=\frac{\tilde H_0}{t} +
\frac{\tilde H_1}{t_0 - t}\ . \ee Here $\tilde H_0$, $\tilde H_1$,
and $t_0$ are positive constants. When $t$ is small, $H$
(\ref{FGB15}) behaves in a way corresponding to the perfect fluid
case, with $w_{eff}=-1 + 2/3\tilde H_0$. Then, if we choose $\tilde
H_0=2/3$, the matter dominated universe occurs. On the other hand,
when $t\sim t_0$ is large, $H$ behaves as in the phantom universe
with $w_{eff}=-1 - 2/3\tilde H_1<-1$ and a big rip singularity at
$t=t_0$ will appear. The three-year WMAP data are analyzed in
Ref.~\cite{Spergel}, which shows that the combined analysis of WMAP
with the supernova Legacy survey (SNLS) constrains the dark energy
equation of state $w_{DE}$ pushing it clearly towards the
cosmological constant value. The marginalized best fit values of the
equation of state parameter at 68$\%$ confidence level are given by
$-1.14\leq w_{DE} \leq -0.93$, which corresponds to $\tilde
H_1>10.7$ as $\tilde H_1$ is positive. In the case when one takes as
a prior that the universe is flat, the combined data gives $-1.06
\leq w_{DE} \leq -0.90 $, which corresponds to $\tilde H_1>25.0$.
Therefore, the possibility that $w_{DE}<-1$ has not been excluded.

Finally, an additional example is the $\Lambda$CDM-type cosmology:
\be \label{LCDM1} g(t)=\frac{2}{3(1+w)}\ln \left[\alpha \sinh
\left(\frac{3(1+w)}{2l}\left(t - t_0 \right)\right)\right]\ ,\quad
\alpha^2\equiv \frac{1}{3} l^2 \rho_0 a_0^{-3(1+w)}\ . \ee Here $l$
is the length scale given by the cosmological constant $l\sim
\left(10^{-33}\,{\rm eV}\right)^{-1}$ and $t_0$ is a constant. The
time-development of the universe given by $g(t)$ (\ref{LCDM1}) can
be realized in the usual Einstein gravity with a cosmological
constant $\Lambda$ and cold dark matter (CDM), which could be
regarded as  dust. The corresponding scalar potentials can indeed be
written explicitly, but they are quite complicated functions.

%%%%%%%%%%%%%%%%%%%%%%%%%%%%%%%%%%%%

\section{Construction of the de-Sitter solution for a
general effective action}

Let us study the possibility of realizing de-Sitter space from the
scalar field equation and the Friedman equation. The coefficients
$c_1$, $c_2$, $c_3$,
 and also $\delta_H$, $\delta_B$ in the action
(\ref{stcor})-(\ref{ccc})
depend on what kind of string
theory, that is, bosonic string, type II superstring theory,
or heterotic string theory, we are considering.
Furthermore, these coefficients could depend on the details of
 compactification. Moreover,
the suitable compactification often induces a scalar potential.
Hence, here we consider the conditions for the coefficients which
allow the de-Sitter space solution. In other words, we assume the
possibility of a more general effective action like in previous
section.

In the de-Sitter space, the Hubble rate $H$ is a constant \be
\label{dS1} H=H_0\ , \ee and all the curvatures are covariantly
constant. We
 also assume the scalar field $\phi$ to be a constant:
\be \label{dS2} \phi=p_0\ . \ee For simplicity,  $c_3$ terms are
neglected and the scalar potential $V(\phi)$ is assumed to be given
by \be \label{dS3} V(\phi)=V_0\e^{-2\phi/\phi_0}\ . \ee Then the
scalar equation has the following form: \be \label{dS4} 0=V_0 +
24c_1 \alpha' x^2 + 96 c_2 {\alpha'}^2 x^3\ ,\ \quad x\equiv
H_0\e^{2p_0/\phi_0}\ . \ee On the other hand, the Friedmann equation
is reduced to \be \label{dS5} 0=V_0 + {3}x - 12 c_2 {\alpha'}^2 x^3\
. \ee By eliminating $c_2$ from (\ref{dS4}) and (\ref{dS5}), one
obtains \be \label{dS6} 0=3V_0 + {8}x + 8c_1\alpha' x^2\ . \ee On
the other hand, by eliminating $V_0$ from (\ref{dS4}) and
(\ref{dS5}),
 we get
\be \label{dS7} 0=-{1} + 8c_1 \alpha' x + 36 {\alpha'}^2 c_2 x^2\ .
\ee And by further eliminating the $x^2$ term from (\ref{dS6}) and
(\ref{dS7}), we find \be \label{dS8} x=x_0\equiv \frac{12\alpha'c_2
V_0 + {c_1}}{8\alpha'\left(c_1^2 - {4\alpha' c_2}\right)}\ . \ee This
expression for $x$ is not always a solution of the two independent
equations (\ref{dS4}) and (\ref{dS5}), or equivalently (\ref{dS6})
and (\ref{dS7}). The condition for $x$ in (\ref{dS7}) to be  a
solution can be obtained by substituting the expression (\ref{dS8})
into (\ref{dS6}
 (or equivalently (\ref{dS7})):
\be \label{dS9} 0=24V_0\alpha'\left(c_1^2 - {4\alpha'c_2}\right)^2 +
{8}\left(12\alpha'c_2 V_0 + {c_1}\right) \left(c_1^2 -
{4\alpha'c_2}\right) + c_1 \left(12\alpha'c_2 V_0 + {c_1}\right)^2\
. \ee In the particular case when $V_0=0$, Eq.~(\ref{dS9}) has the
following
form: \be \label{dS10} 9c_1^2={32\alpha' c_2}\ , \ee and (\ref{dS8})
gives \be \label{dS11} x_0=\frac{c_1}{8\alpha'\left(c_1^2 -
{4\alpha' c_2}\right)}=-\frac{1}{\alpha' c_1}\ . \ee Therefore, if
$c_1<0$, there is a possibility that there could occur a de-Sitter
space solution. We should note that, even if the solution exists, the
Hubble rate $H$  itself cannot be determined uniquely.
In fact, since
\be
\label{dS12}
H=\sqrt{x_0}\e^{p_0/\phi_0}\ ,
\ee
by choosing $p_0$ properly, the value of $H$ itself could be
arbitrarily changed. Then the value of $H$ could be
determined by the initial condition.

Hence, we have presented in the above the condition to be satisfied
by the coefficients of our effective action, which leads naturally
to a de-Sitter universe. In
other words, if this condition is fulfilled, the early-time and (or)
late-time universe can be inflationary (non-singular) due to stringy
effects. It goes without saying that, again, the stability of such
de-Sitter universe should be checked in each case, as it was done
for the bosonic string earlier.

%%%%%%%%%%%%%%%%%%%%%%%%%%%%%%%%%%%%%

\section{Conclusions}

In this paper we have considered string loop corrections to the
Einstein-Hilbert action given by (\ref{stcor}), with a dynamical
dilaton $\phi$. We have explored the cosmological dynamics of the
corresponding
modified gravity in the framework of a low-energy string effective
action. For simplicity, we have ignored the contribution of the
background (radiation/matter) energy density. The higher-order
string corrections to gravity, specially the third-order correction
in $\alpha'$, crucially depend upon the string type. The evolution
equations are quite involved and it is difficult to analyze them in
general. Taking a different route, we have conjectured a particular
solution, $H=h_0/t$, $\phi=\phi_0\ln t/t_1$ for
$h_0>0$[$H=h_0/(t_s-t)$, $\phi=\phi_0\ln(t_s-t)/t_1$ when $h_0<0$].
This solution is important from the dark-energy viewpoint; we have
carefully checked its viability by enforcing a consistency check on the
parameters of the solution. This consistency requirement constraints
the range of the parameter $h_0$, which defines the effective
equation of state.

The model based upon a Type II string turned out to be the simplest
to investigate semi-analytically. The possible range of  $h_0$ in
this case is given by $0<h_0<1/3$ corresponding to an uninteresting
equation of state ($w_{eff}>1$). Type II is clearly ruled out
because the string expansion parameter cannot be negative.
Nevertheless, the situation might be improved in presence of
matter. It would also be interesting to examine the fate of a phantom
universe in presence of higher curvature corrections with a massless
dilaton.

In the case of the heterotic string, the dark energy solution exists
for $h_0$ varying as $0.8<h_0<5.04$, which corresponds to
$w_{eff}\ge -0.868$. WMAP$_3$ data analyzed with the SNLS survey
constraints the dark energy equation of state as
$w_{DE}=-0.97^{+0.07}_{+0.09}$ at the $68\%$ confidence level, which
puts the heterotic string model under pressure. However, the
combined (CMB+SNLS+LSS) data forces the dark energy equation of
state parameter to vary as $-1.001<w_{DE}<-0.875$. Thus, the
heterotic string is only marginally compatible with observations.

Cosmological dynamics based upon the bosonic string turn out to be
distinguished amongst all possible string types. Indeed, the
consistency of the model leads to an effective equation of state
given by $-1\leq w_{eff} \leq -0.17$, which is clearly compatible
with data. The stability analysis is difficult to carry out in this
case, in general. We have studied the de-Sitter case ($h_0\to
\infty$) in the bosonic case separately and demonstrated its
stability.

A more general action, with higher-curvature corrections coefficients
depending on a dilaton was also considered. The general reconstruction
method could be developed for such theory, so that a realistic universe
expansion history can be obtained within some class of scalar
potentials. An example which proposes a matter-dominance era before
cosmic acceleration (quintessence, phantom era or  $\Lambda CDM$
cosmology) is presented. The de-Sitter universe in such a general
theory (as well as for the bosonic string) can arise quite naturally.
It is known that, with the addition of a scalar-Gauss-Bonnet term
only, the low-energy string effective action can indeed help
in the resolution of the
initial singularity problem \cite{ant}. The appearance of de-Sitter
solution in the general case with higher curvature corrections clearly
indicates that the resolution of an initial and/or a final
singularity of any type (for the classification of future,
finite-time singularities, see \cite{tsujikawa}) could be possible
taking into account higher-order string loops.

We conclude that it is not so easy to get the natural dark energy
universe from a low-energy string effective action (at the very least
up to $R^4$ corrections). It could turn out that use of even
higher order terms is necessary, or that the consideration of a
different compactification might lead to a more realistic universe.
In this respect, it could be expected that taking into account
stringy non-perturbative effects might help. For instance, one future
possibility is to consider not only higher curvature corrections but
also to include negative powers of such terms (an inverse $\alpha'$
expansion?) like in the models with positive and negative powers of the
curvature\cite{NO}.

%%%%%%%%%%%%%%%%%%%%%%%%%%%%%%%%%%%%%
\section*{ACKNOWLEDGEMENTS} We
thank A.~Toporensky, P~V.~Tretjakov and S.~Tsujikawa for fruitful
discussions. MS is supported by JSPS grant No FY2007 and by ICTP and
IUCAA  through their associateship program. The work of IT is
supported by ICCR fellowship. EE and SDO have been supported in part
by MEC (Spain), projects BFM2006-02842 and FIS2005-25313-E, and by
AGAUR (Gene\-ra\-litat de Catalunya), contract 2005SGR-00790. SN has
been supported in part by Monbusho under grant no.18549001 (Japan)
and 21th Century COE Program of Nagoya University provided by JSPS
(15COEG01).
%%%%%%%%%%%%%%%%%%%%%%%%%%%%%%%%%%%%%

\section*{APPENDIX}

\section*{General evolution equations}
The general evolution equations are obtained by varying the action
(\ref{eq:action}) with respect to $\phi$ and the lapse function $N$.
The scalar field equation takes the following form
\begin{eqnarray}\label{GFeqn}
0&=&{\ddot \phi} +3 H \dot \phi \nonumber \\
&&+ \frac{1}{\phi_0} \left \lbrace -48c_1\alpha 'e^{2\phi/\phi_0}
H^4 -192c_2{\alpha '}^2e^{4\phi/\phi_0} H^6 +c_3\left
[432\zeta(3)-2736\delta_B-432\delta_H \right ]{\alpha
'}^3e^{6\phi/\phi_0} H^8  \right. \nonumber \\ &&\left.-48c_1\alpha
'e^{2\phi/\phi_0} H^2{\dot H}-288c_2{\alpha '}^2e^{4\phi/\phi_0}
H^4{\dot H} +c_3\left [864\zeta(3)-5472\delta_B-864\delta_H\right
]{\alpha '}^3e^{6\phi/\phi_0} H^6{\dot H}
\right.\nonumber\\
&&\left.-288c_2{\alpha '}^2e^{4\phi/\phi_0} H^2{\dot H}^2 +c_3\left
[792\zeta(3)-7056\delta_B+1152\delta_H\right ]{\alpha
'}^3e^{6\phi/\phi_0} H^4{\dot H}^2 -96c_2{\alpha
'}^2e^{4\phi/\phi_0}{\dot H}^3
\right.\nonumber\\
&&+c_3\left.[288\zeta(3)-4320\delta_B-720\delta_H]{\alpha
'}^3e^{6\phi/\phi_0} H^2{\dot H}^3 +
c_3[36\zeta(3)-1080\delta_B-180\delta_H]{\alpha '}^3e^{6\phi/\phi_0}
{\dot H}^4 \right\rbrace.
\end{eqnarray}
The Friedman equation in the general case is given by
\begin{eqnarray}\label{GFfr}
3 H^2 &=& \frac{1}{2}{\dot \phi}^2 +24c_2{\alpha
'}^2e^{4\phi/\phi_0} H^6 -c_3\left
[72\zeta(3)-456\delta_B-72\delta_H\right ]{\alpha
'}^3e^{6\phi/\phi_0} H^8 -432c_2{\alpha '}^2e^{4\phi/\phi_0}
H^4{\dot H}
\nonumber\\
&&+c_3\left [792\zeta(3)-7056\delta_B-1152\delta_H\right ]{\alpha
'}^3e^{6\phi/\phi_0} H^6{\dot H} -288c_2{\alpha '}^2e^{4\phi/\phi_0}
H^2{\dot H}^2
\nonumber\\
&&+c_3\left [828\zeta(3)-10008\delta_B-1656\delta_H\right ]{\alpha
'}^3e^{6\phi/\phi_0} H^4{\dot H}^2 +48c_2{\alpha
'}^2e^{4\phi/\phi_0} {\dot H}^3
\nonumber\\
&& +c_3\left [168\zeta(3)-3600\delta_B-600\delta_H\right ]{\alpha
'}^3e^{6\phi/\phi_0} H^2{\dot H}^3 -c_3\left
[18\zeta(3)-540\delta_B-90\delta_H\right ]{\alpha
'}^3e^{6\phi/\phi_0} {\dot H}^4
\nonumber\\
&& -144c_2{\alpha '}^2e^{4\phi/\phi_0} H^3{\ddot H}^3 +c_3\left
[264\zeta(3)-2352\delta_B-384\delta_H\right ]{\alpha
'}^3e^{6\phi/\phi_0} H^5{\ddot H} -144c_2{\alpha
'}^2e^{4\phi/\phi_0} H{\dot H}{\ddot H}
\nonumber\\
&& +c_3\left [288\zeta(3)-4320\delta_B-720\delta_H\right ]{\alpha
'}^3e^{6\phi/\phi_0} H^3{\dot H}{\ddot H} +c_3\left
[72\zeta(3)-2160\delta_B-360\delta_H\right ]{\alpha
'}^3e^{6\phi/\phi_0} H{\dot H}^2{\ddot H}
\nonumber\\
&& + \frac{\dot \phi}{\phi_0} \left \lbrace -48c_1\alpha
'e^{2\phi/\phi_0} H^3 -288c_2{\alpha '}^2e^{4\phi/\phi_0} H^5
+c_3\left [864\zeta(3)-5472\delta_B-864\delta_H\right ]{\alpha
'}^3e^{6\phi/\phi_0} H^7  \right. \nonumber \\
&&\left.-576c_2{\alpha '}^2e^{4\phi/\phi_0} H^3{\dot H} +c_3\left
[1584\zeta(3)-14112\delta_B-2304\delta_H\right ]{\alpha
'}^3e^{6\phi/\phi_0} H^5{\dot H}
\right.\nonumber\\
&&\left. -288c_2{\alpha '}^2e^{4\phi/\phi_0} H{\dot H}^2 +c_3\left
[864\zeta(3)-12960\delta_B-2160\delta_H\right ]{\alpha
'}^3e^{6\phi/\phi_0} H^3{\dot H}^2
\right. \nonumber\\
&&\left. +c_3\left [144\zeta(3)-4320\delta_B-720\delta_H\right
]{\alpha '}^3e^{6\phi/\phi_0} H{\dot H}^3 \right\rbrace.
\end{eqnarray}

\end{document}